\documentclass{article} %
\usepackage{iclr2026_conference,times}
\iclrfinalcopy % de-anonymize; no review header

\usepackage{amsmath,amsfonts,bm}

\def\eqref#1{equation~\ref{#1}}

\def\1{\bm{1}}

\DeclareMathAlphabet{\mathsfit}{\encodingdefault}{\sfdefault}{m}{sl}
\SetMathAlphabet{\mathsfit}{bold}{\encodingdefault}{\sfdefault}{bx}{n}

\usepackage{hyperref}
\usepackage{graphicx}
\usepackage{url}
\usepackage{xspace}
\usepackage{booktabs}
\usepackage{tabularx}
\usepackage{multirow}
\usepackage{array}
\usepackage{xcolor}
\usepackage[normalem]{ulem}
\usepackage{pifont}
\usepackage{makecell}
\usepackage{wrapfig}
\usepackage[inline]{enumitem}
\usepackage{pgfplots}
\usepgfplotslibrary{groupplots}
\pgfplotsset{compat=1.18}
\usepackage{float}

\newcommand{\benchmark}{\text{SRE-Bench}}
\newcommand{\experts}{\text{RE experts}}

\newif\ifrobinreview
\ifdefined\robinaccepted
  \robinreviewfalse
\else
  \robinreviewtrue
\fi
\newcommand{\robinremoved}[1]{\ifrobinreview{\color{gray}\sout{#1}}\fi}

\newcolumntype{Y}{>{\raggedright\arraybackslash}X}

\newcommand{\cmark}{\textcolor{green!60!black}{\ding{51}}}
\newcommand{\xmark}{\textcolor{red!80!black}{\ding{55}}}
\newcommand{\pmark}{\textcolor{orange!85!black}{\ensuremath{\triangle}}}

\newcommand{\benchmarkLoC}{\text{16{,}915.8}}

\newcommand{\domainicon}[1]{\raisebox{-0.1em}{\includegraphics[height=1em]{figures/#1_icon.pdf}}\,}

\newcommand{\eg}{\textit{e.g.},}
\newcommand{\ie}{\textit{i.e.},}

\title{The Next Challenge for Agentic Cybersecurity: A Realistic, Contamination-Free Reverse Engineering Benchmark}

\author{Jeremy Spence$^{1\dagger}$\quad Nicholas Assaderaghi$^{1}$\quad Feng Xiao$^{2}$\quad Jinhao Zhu$^{3}$ \\
\bf Nikil Ravi$^{8}$\quad Xiangyu Qi$^{4}$\quad Matthew Jagielski$^{5}$\quad Raluca Ada Popa$^{2,3}$ \\
\bf Eric Wallace$^{4}$\quad Guannan Wei$^{6}$\quad Yangruibo Ding$^{7}$\quad Zhuo Zhang$^{1}$\textsuperscript{\ding{41}} \\[4pt]
\normalfont $^{1}$Columbia University\quad $^{2}$Google DeepMind\quad $^{3}$UC Berkeley\quad $^{4}$OpenAI \\
\normalfont $^{5}$Anthropic\quad $^{6}$Tufts University\quad $^{7}$UCLA\quad $^{8}$Vals AI \\[2pt]
\normalfont $^{\dagger}$\texttt{js6721@columbia.edu}\quad \ding{41}\,\texttt{zz@cs.columbia.edu}
}

\begin{document}

\maketitle

\begin{abstract}
AI agents are rapidly improving in cybersecurity capabilities when the \textit{source code} is available for analysis, yet much of the software most consequential to cybersecurity, including malware, firmware, and proprietary applications, is available only as \textit{binaries}. 
Analyzing such software requires \textit{reverse engineering} (RE): recovering program semantics before the analysis can be meaningfully performed. 
However, evaluating agentic RE poses a fundamental challenge: 
realistic and useful benchmark instances must 
(1) be absent from
LLMs' training data to prevent models from taking shortcuts by memorization, and 
(2) represent the scale and anti-analysis protections of real-world binaries.
To faithfully evaluate the RE capability of frontier models,
we introduce \textbf{\benchmark{}}, the first realistic, contamination-free RE benchmark.
Built entirely from scratch by \experts{} with over 5,000 expert hours, \benchmark{} comprises 19 private, real-world-scale programs averaging 16.9K lines of code. 
We further developed 44 in-house anti-analysis mechanisms, yielding 262 binary instances and 1,572 deterministically graded tasks.
We evaluated 13 agentic settings across 11 models: eight in public-facing settings and five in internal unconstrained settings with cyber safeguards disabled and no budget cap.
Our results show that 
realistic RE remains challenging for frontier agents: GPT-5.6-Sol and Claude-Fable-5.1, despite their strong source-code cybersecurity capabilities, fully solve only 31.5\% and 26.9\% of graded instances, respectively. 
This result suggests that success in source-code security does not necessarily translate into effective binary analysis. 
Without a budget cap and safety guard,
GPT-6-Astra achieves a near-perfect pass@4 score;
however, reliably identifying the correct candidate still poses a significant challenge.
Our analysis further reveals that agents 
are relatively insensitive to compiler optimization and static linking, and 
controlled ablations confirm that both contamination control and realistic scale are essential to understand agents' RE capability. %
Together, these findings highlight RE as a distinct frontier for agentic cybersecurity and establish \benchmark{} as a rigorous testbed for measuring progress.
\looseness=-1

\end{abstract}

\section{Introduction}
\label{sec:intro}

Much of the software most consequential to cybersecurity reaches analysts only as \textit{binaries, without source code}. 
This is true at both sides of the threat landscape. 
On the defensive side, the high-value systems (\eg{} proprietary enterprise software, security appliances, and firmware) are frequent attack targets but are typically distributed only in binary form.
For example, according to the authoritative catalog~\citep{cisa2026kev}, 46.5\% of vulnerabilities exploited in the wild come from vendors that do not release source code.
Google similarly reports that more than 48\% of the zero-days in 2025 targeted proprietary enterprise software~\citep{gtig2025zeroday}. 
On the offensive side, attackers deliberately distribute malicious payloads as obfuscated binaries to impede analysis, with approximately 732{,}000 new malicious samples reported each day~\citep{vanliebergen2023virustotal}. Anthropic also makes the distinction between binary exploitation and source exploitation as two distinct capabilities~\citep{anthropic2026opus55}.
As autonomous cybersecurity agents advance, binary software becomes an essential and urgent evaluation target to understand the state-of-the-art model capability.

Unlike structured source code, binary code is represented as raw bytes and is not directly intelligible to analysts.
Over the past several decades, cybersecurity researchers have developed \textit{reverse engineering} (RE)~\citep{shoshitaishvili2016sok, lee2011tie, balakrishnan2005wysinwyx, song2008bitblaze}
to recover high-level program semantics from this opaque representation. 
RE now forms the foundation of binary-first security analysis: before analysts can reason about vulnerabilities, patches, or exploits, they must first determine what a binary program does.
Notably, RE requires additional capabilities beyond those %
required in downstream cybersecurity tasks~\citep{mantovani2022re, votipka2020observational}; 
it entails interpreting low-level machine semantics, inferring intent from incomplete evidence, and, in harder cases, overcoming adversarial obfuscation and packing that 
intentionally obscure the binary code.
To extend agentic cybersecurity beyond high-level source code, AI agents must therefore learn how to tackle RE and be evaluated on it as a distinct capability.

However, constructing an RE benchmark that faithfully reflects real-world practice is challenging.

\textbf{An RE benchmark is valid only if its targets are unknown to the model.}
The goal of RE is to recover the semantics of {\textit{unknown binaries}} (\eg{} malware or proprietary software).
Analysts therefore begin with no prior knowledge of the target.
Benchmarks built from publicly available source code break this premise,
because the target may already appear in pretraining data and be recognizable during evaluation~\citep{al2024traces,jain2025livecodebench,openai2026why}.
Once a model recognizes the target, its prior knowledge (\eg{} the target's high-level purpose) can help bypass much of the program-understanding process.
Note that this risk differs from contamination in many other cybersecurity tasks, where only exact, fine-grained leakage is concerning~\citep{ding2024vulnerability}.
In RE, even a coarse degree of leakage is damaging~\citep{siegmund2016program}, as it provides crucial, top-down guidance for program interpretation~\citep{biggerstaff1993concept}.
Our empirical results in \S\ref{sec:evaluation} further confirm that such data contamination can inflate RE task performance.

\textbf{An RE benchmark must also match the scale and protection of real-world targets.}
Recent work shows that cybersecurity benchmarks built from CTF-style challenges or toy programs fail to predict agent performance on real-world 
targets~\citep{wang2026exploitgym,zhang2026bountybench}.
Evaluating RE tasks can easily fall into the same pitfall. %
The complexity of binary analysis arises from two sources: 
1) the behavior of \textit{target program} and 
2) the \textit{protection layer} that gates access and obscures understanding of proprietary software.
At the program level, real-world software can span thousands, if not hundreds of thousands, of lines of code.
Because RE is fundamentally a program-understanding task, 
the combinatorial complexity of program behaviors
grows exponentially with the scale~\citep{wettel2011software};
capability measured on toy programs therefore does not extrapolate~\citep{rugaber1995program}.
At the protection level, high-value targets are often wrapped in multiple anti-analysis layers, such as sophisticated and sometimes bespoke obfuscation~\citep{cheng2021obfuscation}.
A faithful benchmark must therefore pair real-world-scale programs with protection beyond textbook schemes.

Together, these requirements rule out some straightforward construction
strategies one may expect: 
Using open-source projects provides real-world artifacts, yet to control
contamination, we must exclude them.
Applying local modifications to such projects is also insufficient, 
because their high-level architecture may remain recognizable.
As a result, to faithfully evaluate RE capability of frontier models,
we must construct binary programs with real-world complexity entirely from scratch,
albeit it can be highly time-consuming and expertise-intensive.
To this end, 
we have invested over 5,000 domain-expert hours in developing a clean-room RE
benchmark. 
To our knowledge, no prior RE benchmark has demanded a comparable investment of
expert effort.

\textbf{\benchmark{}: The First Realistic, Contamination-Free RE Benchmark.}
We developed 19 in-house programs, averaging more than \benchmarkLoC{} lines of code (LoC), across five representative RE domains: \text{network protocols}, \text{firmware}, \text{games}, \text{format parsers}, and \text{malware}. 
These domains correspond to the major categories identified by a recent survey of reverse-engineering discussions on Stack Exchange~\citep{kabir2026restack}.
We also built a comprehensive anti-analysis suite with 44 state-of-the-art protection primitives, over half of which have no public implementations.
The suite itself required over 27K LoC to implement.
By pairing these private programs with in-house protection settings, \benchmark{} yields \textbf{262 contamination-free instances with realistic difficulty}.
Each instance defines six deterministically verifiable tasks. %
In total, \benchmark{} provides 1,572 RE tasks.

\textbf{Experimental Results Reveal Persistent Challenges in Agentic RE.} We evaluate frontier agentic LLMs under both standardized public and model-provider-internal configurations. Most evaluated agents struggle with realistic RE: despite their strong source-code security capabilities~\citep{openai2026gpt56, anthropic2026mythos5}, GPT-5.6-Sol and Claude-Fable-5.1 fully solve only 31.5\% and 26.9\% of graded instances, respectively, in the standardized evaluation. These results suggest that success in source-code security does not necessarily translate into effective binary analysis. GPT-6-Astra demonstrates substantial progress, achieving 99.2\% pass@4 under the model-provider-internal evaluation, but selecting or validating the correct solution still remains an important open challenge. 

Our in-house protections further expose substantial differences in binary-analysis capability: with the build held fixed, they reduce GPT-5.6-Sol’s mean score from 4.69 to 2.50 out of 6 and drive weaker models to near zero. GPT-6-Astra largely maintains its score, but incurs 1.5$\times$ the monetary cost per point on protected binaries while weaker models incur up to 35.3$\times$ cost. The failure modes also differ from those of human analysts. Optimization and linking, the classical obstacles of manual RE, barely register, whereas stripping symbols is costly, suggesting that current agents lean on lexical anchors more than on instruction-level reasoning. Together, these findings establish SRE-Bench as a testbed for advancing both binary understanding and reliable solution validation.

\textbf{Our Contributions.}
This work makes four contributions.
First, we identify RE as the next frontier for agentic cybersecurity capability, and argue that the field must move beyond source-code benchmarks to measure agents on binary-first workflows.
Second, we show empirically that both program complexity and data contamination substantially affect RE evaluation, establishing them as essential design criteria %
rather than optional refinements.
Third, we present \benchmark{}, the first RE benchmark that addresses both requirements: 262 instances built from scratch at the real-world scale (averaging \benchmarkLoC{} LoC), paired with a 27K-LoC anti-analysis suite implementing 44 in-house protection primitives. %
Lastly, we give the first systematic characterization of agentic RE capability in a real-world setting with \benchmark{}. Our evaluation on state-of-the-art AI agents reveals that strong source-code security capabilities do not necessarily translate into effective binary analysis.

\section{Related Work}
\label{sec:related}

\newcommand{\protEasy}{\textcolor{red!75!black}{\textbf{Easy}}}
\newcommand{\protTextbook}{\textcolor{orange!85!black}{\textbf{Textbook}}}
\newcommand{\protRealWorld}{\textcolor{green!45!black}{\textbf{Real-World}}}

\begin{table*}[t]
\centering
\scriptsize
\setlength{\tabcolsep}{3.2pt}
\renewcommand{\arraystretch}{1.08}
\caption{
    Comparing \benchmark{} with existing, end-to-end RE benchmarks for AI agents.
    \textbf{\# Instances} denotes the number of RE instances.
    \textbf{RE-Centric} indicates whether the benchmark is primarily designed for RE.
    \textbf{Cont. Ctrl.} denotes control for program-identity contamination.
    \textbf{Prog. Cplx.} summarizes program complexity, and \textbf{LoC} reports the average lines of code per instance for non-CTF benchmarks.
    \textbf{Prot. Cplx.} summarizes the complexity of binary-protection mechanisms.
}
\label{tab:agentic-re-benchmarks}
\vspace{2pt}
\begin{tabular}{@{}lcccccc@{}}
\toprule
\textbf{Benchmark} &
\textbf{\# Instances} &
\textbf{RE-Centric} &
\textbf{Cont. Ctrl.} &
\textbf{Prog. Cplx.} &
\textbf{LoC} &
\textbf{Prot. Cplx.} \\
\midrule

NYU CTF Bench~\citep{shao2024nyuctfbench} &
51 & \xmark & \xmark & Existing CTF & -- & \protEasy \\

Cybench~\citep{zhang2025cybench} &
6 (15 \textit{tasks}) & \xmark & \xmark & Existing CTF & -- & \protEasy \\

CTF-Dojo~\citep{zhuo2025training} &
123 & \xmark & \xmark & Existing CTF & -- & \protEasy \\

CTFTiny~\citep{shao2026effectiveoffensivellmagents} &
16 & \xmark & \xmark & Existing CTF & -- & \protEasy \\

CREBench~\citep{chen2026crebench} &
432 (1728 \textit{tasks}) & \cmark & \pmark & Toy Program & 526.2 & \protEasy \\

AgentRE-Bench~\citep{agentrebench2026} &
13 (70 \textit{tasks}) & \cmark & \cmark & Toy Program & 63.5 & \protTextbook \\

CrackMeBench~\citep{david2026crackmebench} &
12 & \cmark & \cmark & Toy Program & 36.1 & \protTextbook \\

\midrule
\benchmark{} (Ours) &
262 (1572 \textit{tasks}) & \cmark & \cmark &
\textbf{Real-World Software} &
\textbf{\benchmarkLoC{}} &
\protRealWorld \\

\bottomrule
\end{tabular}
\renewcommand{\arraystretch}{1.00}
\vspace{-5pt}
\end{table*}

\textbf{Source-Code Security Benchmarks for AI Agents.}
A broad line of work evaluates AI agents on source-code security tasks.
\citet{chen2023diversevul, liu2024vuldetectbench, yildiz2025benchmarking} introduce vulnerability-detection benchmarks that measure whether models can identify flaws across entire codebases.
\citet{wang2025cybergym, lee2026sec, pu2026patch} introduce PoC-generation benchmarks that evaluate whether agents can reproduce vulnerabilities using concrete inputs.
\citet{zhu2025cve, wang2026exploitgym, lee2026exploitbench} further propose exploit-development benchmarks that assess whether agents can construct working exploits.
Other security-oriented software-engineering benchmarks evaluate complementary code-security workflows~\citep{nie2026secodeplt,lee2026sec,lee2026secpro,wang2025cybergym}.
\benchmark{} instead targets the missing bridge to binaries,
testing whether AI agents' success on source code extends to broader security workflows.

\textbf{Traditional RE Benchmarks.}
Many RE benchmarks evaluate specific intermediate artifacts, such as function and variable names~\citep{koller2026reforge}, types~\citep{won2026rebench,soni2025benchmarking}, and decompiled source code~\citep{tian2026decompile,gao2025decompilebench}.
However, artifact-level accuracy does not fully capture end-to-end program understanding: an agent may correctly infer program behavior or identify vulnerabilities without recovering the original names, types, or source code.

\textbf{Binary-Based Software Engineering Benchmarks.}
ProgramBench~\citep{yang2026programbench} gives agents the compiled binary of public open-source projects and their documentation and asks them to reimplement the program's behavior without decompilers ~\citep{anthropic2026opus55,anthropic2026sonnet5}. The targets are well-known open-source programs (\eg{} \texttt{jq}, \texttt{SQLite}, \texttt{FFmpeg}) whose purpose is disclosed through provided documentation. This measures behavioral reimplementation rather than
semantic recovery based purely on the binary, which is what SRE-Bench targets.

\textbf{End-to-End  RE Benchmarks for AI Agents.}
Closer to our setting, several benchmarks evaluate AI agents on RE tasks, but they still diverge sharply from real-world practice, as summarized in \autoref{tab:agentic-re-benchmarks}.
Among them, CTF-derived benchmarks~\citep{shao2024nyuctfbench, zhang2025cybench, zhuo2025training, shao2026effectiveoffensivellmagents} include RE tasks, but only as one subcategory.
Because these benchmarks are drawn from existing CTF competitions whose intended solutions are often publicly available, they carry substantial risk of data contamination.
Moreover, most of their collected challenges are entry-level, 
which fail to capture real-world complexity in either target programs or binary protections.

A second line of work focuses specifically on RE tasks and constructs benchmark programs from scratch to reduce data-contamination risk.
However, these benchmarks are limited to toy programs.
CREBench~\citep{chen2026crebench} targets crypto-oriented RE and contains 432 instances generated from 48 small programs, averaging only 526.2 LoC per program.
Meanwhile, it still carries contamination risk, as 39 of its 48 programs are slight modifications of open-source projects.
Its obfuscation is rudimentary as well, limited to naive string-XOR obfuscation.
AgentRE-Bench~\citep{agentrebench2026} and CrackMeBench~\citep{david2026crackmebench} provide stronger contamination control by using 13 and 12 self-developed programs, respectively. 
However, they are even smaller in scale, averaging only 63.5 and 36.1 LoC per instance. 
Their binary protections also remain textbook-level, such as simple keygens and miniature virtual machines. 
For example, AgentRE-Bench implements all protection logic in only 376 lines of C code.
In contrast, \benchmark{} is designed to evaluate agentic RE under substantially more realistic conditions while maintaining strict contamination control. 
Its programs are developed from scratch at real-world scale, averaging \benchmarkLoC{} LoC per instance. 
Its protection layer is a comprehensive anti-analysis suite built with over 27K LoC, incorporating many techniques widely used in commercial obfuscators but lacking mature public implementations.

\section{Benchmark}
\label{sec:benchmark}

\begin{figure*}[t]
    \centering
    \includegraphics[width=0.9\textwidth]{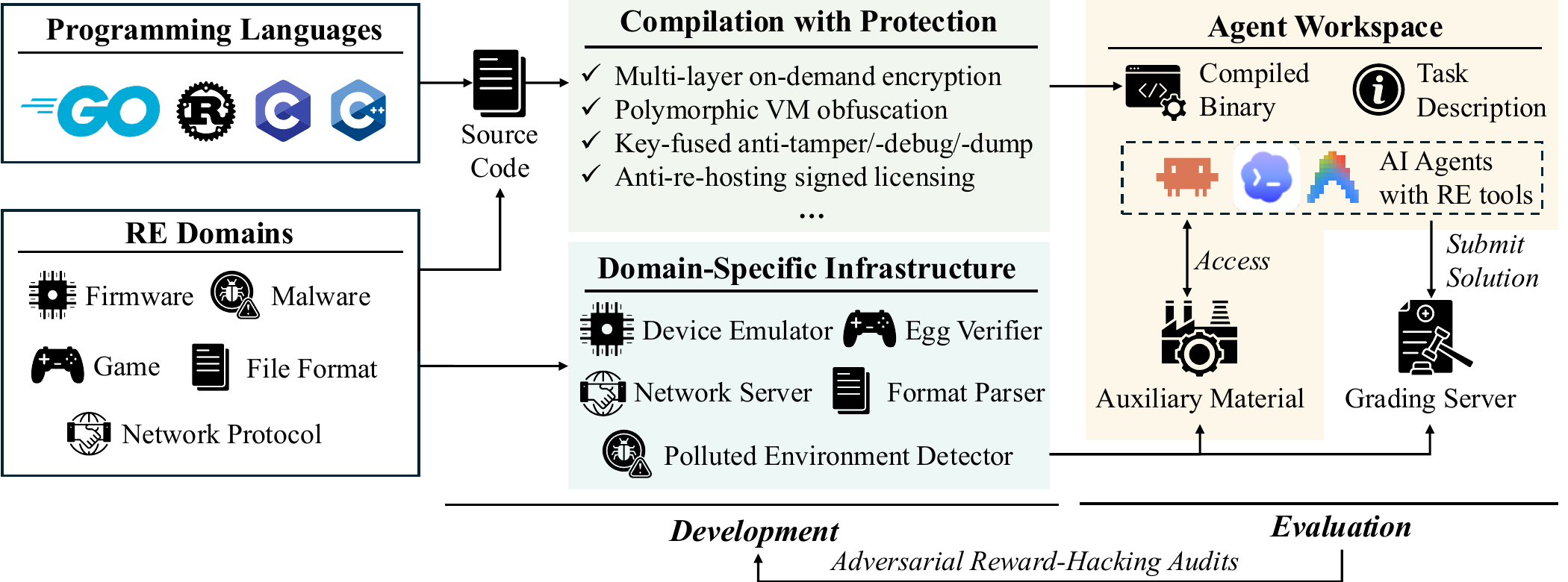}
    \caption{
        Overview of the \benchmark{} construction and evaluation pipeline.
        \experts{} develop each program and its evaluation infrastructure from scratch against a private specification.
        Programs are compiled into diverse build variants and optionally hardened by our in-house protection suite, yielding contamination-free binary instances.
        At evaluation time, the agent receives a binary plus domain-specific auxiliary materials, then submits a solution to a deterministic grader.
        The same evaluation pipeline is reused during development to run adversarial reward-hacking audits (Opus 4.8 and GPT-6-Astra), whose findings drive iterative hardening of both the programs and infrastructure.
    }
    \vspace{-12pt}
    \label{fig:benchmark_pipeline}
\end{figure*}

\autoref{fig:benchmark_pipeline} illustrates the construction pipeline of \benchmark{}.
The benchmark is built entirely in-house: \experts{}
develop the target programs and their evaluation infrastructure from scratch, compile and optionally harden them into diverse binary instances, and score every submission with a deterministic grader.
\textit{Adversarial reward-hacking audits} further close the loop by pitting agents against in-progress instances under the real grader and patching every shortcut that earns credit without requiring genuine RE.
We deliberately confine \benchmark{} to reverse engineering and exclude downstream cybersecurity tasks such as exploit development, for two reasons: 1) including such tasks would conflate an agent's RE ability with its separate cybersecurity skill, and 2) reverse engineering is a distinct capability that merits measurement on its own~\citep{mantovani2022re, votipka2020observational}.
The following subsections detail the RE domains and the protection suite.

\begin{wraptable}[11]{r}{0.48\textwidth}
\centering
\vspace{-2em}
\scriptsize
\setlength{\tabcolsep}{4pt}
\renewcommand{\arraystretch}{1.08}
\caption{
    LoC per program (domain$\times$language).
    Firmware omits Go, whose runtime makes it unsuitable for bare-metal embedded targets.
}
\label{tab:benchmark_loc}
\begin{tabular}{@{}lrrrr r@{}}
\toprule
\textbf{Domain} & \textbf{C} & \textbf{C++} & \textbf{Rust} & \textbf{Go} & \textbf{Total} \\
\midrule
Network Protocol & 17,513 & 15,754 & 18,052 & 29,346 & 80,665 \\
Game             & 19,922 & 19,208 & 19,692 & 22,873 & 81,695 \\
File Format      & 16,894 & 12,291 & 12,732 & 21,144 & 63,061 \\
Malware          & 14,610 & 12,972 & 16,726 & 22,897 & 67,205 \\
Firmware         & 8,874  & 8,892  & 11,009 & --     & 28,775 \\
\midrule
\textbf{Total} & 77,813 & 69,117 & 78,211 & 96,260 & 321,401 \\
\bottomrule
\end{tabular}
\renewcommand{\arraystretch}{1.00}
\end{wraptable}

\textbf{Development Methodology.}
All artifacts (\ie{} the target programs, the evaluation infrastructure, and the protection suite) were developed by \experts{} with an average of six years of experience.
Each program was independently implemented from scratch based on a private design specification that is never released. 
The 19 programs span four implementation languages (\ie{} C, C++, Rust, and Go), follow language-idiomatic programming patterns, and use distinct library stacks (\eg{} \href{https://invisible-island.net/ncurses/announce.html}{\texttt{ncurses}} vs.\ \href{https://github.com/gdamore/tcell}{\texttt{tcell}} vs.\ \href{https://ratatui.rs/}{\texttt{ratatui}}).
No two programs share source code, and none is derived from a public project.
Every program ships with a private reference solution that scores a perfect $6/6$, which serves as both a solvability proof and a regression gate.
We also run iterative adversarial reward-hacking audits: reusing the evaluation pipeline, we pit agents against in-progress instances to surface shortcuts.
For example, an early build retained the internal name of a hidden behavior (\ie{} the target of an RE task) as a readable string in the binary; an agent therefore located the corresponding trigger through string scanning without doing any actual RE.
We removed the string and added a regression check that fails the build on any similar leak.
In total, \benchmark{} comprises over 320K lines of code (\autoref{tab:benchmark_loc}), averaging \benchmarkLoC{} LoC per program; it is 30-470$\times$ larger than prior RE benchmarks.

\textbf{Benchmark Domains.}
\benchmark{} contains 262 instances, spanning five domains that together cover the breadth of real-world RE practice~\citep{kabir2026restack, eilam2011reversing}: \textit{network protocol}, \textit{game}, \textit{format parser}, \textit{malware}, and \textit{firmware}.
\autoref{tab:benchmark_domains} summarizes the characteristics of each domain. More detailed domain descriptions can be found in \autoref{app:benchmark_details}.

\begin{table*}[t]
\centering
\scriptsize
\setlength{\tabcolsep}{3pt}
\renewcommand{\arraystretch}{1.2}
\caption{
    Overview of RE domains.
    \textbf{Target} is what the program does. 
    \textbf{RE Task} is what the agent must accomplish.
    \textbf{Infrastructure} is the component that makes evaluation deterministic.
    \textbf{Auxiliary} is what the agent receives beyond the binary.
    \textbf{\# Programs} reports the number of unique programs in each domain; their implementation languages and Loc can be found in \autoref{tab:benchmark_loc}.
}
\label{tab:benchmark_domains}
\vspace{2pt}
\begin{tabular}{>{\raggedright\arraybackslash}p{1.5cm} >{\raggedright\arraybackslash}p{2.8cm} >{\raggedright\arraybackslash}p{2.8cm} >{\raggedright\arraybackslash}p{2.7cm} >{\raggedright\arraybackslash}p{1.6cm} >{\centering\arraybackslash}p{1.4cm}@{}}
\toprule
\textbf{Domain} &
\textbf{Target} &
\textbf{RE Task} &
\textbf{Infrastructure} &
\textbf{Auxiliary} &
\textbf{\# Programs} \\
\midrule

\domainicon{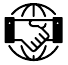}Network Protocol &
A client-server stack for a proprietary, encrypted network protocol &
Recover the wire format and drive a client through the full protocol state machine &
Server that scores the protocol's state-machine coverage of the client &
5 partial packet captures &
4 \\

\midrule
\domainicon{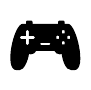}Game &
A playable 20-floor terminal roguelike dungeon crawler &
Trigger hidden behaviors (Easter eggs) unreachable through normal play &
An egg verifier that replays the agent's action trace &
Player manual &
4 \\

\midrule
\domainicon{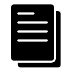}File Format &
A file/directory compressor producing a proprietary archive format &
Reverse the encoder and decode the provided archives byte-exactly &
A held-out decoder that round-trip verifies the developed encoder &
6 challenge files and corresponding passwords &
4 \\

\midrule
\domainicon{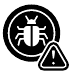}Malware &
A synthetic implant that exhibits malicious behavior without causing real harm &
Reverse the implant's effects while preserving benign user data &
A sandboxed infection-and-cleanup grader &
-- &
4 \\

\midrule
\domainicon{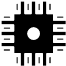}Firmware &
Bare-metal firmware for a locked-down secure microcontroller &
Take progressively fuller control of the device through its interface &
A held-out emulator that models the chip and its peripherals &
Serial console on a remotely hosted emulator &
3 \\

\bottomrule
\end{tabular}
\renewcommand{\arraystretch}{1.00}
\vspace{-15pt}
\end{table*}

\textbf{Protection and Obfuscation}
High-value RE targets are rarely presented as clean, unprotected binaries.
To model this, \benchmark{} includes an in-house binary protection suite that transforms an unprotected Linux ELF into a hardened, functionally equivalent variant.
The suite is implemented entirely from scratch, comprising over 27K lines of Python, C, and assembly, which keeps the protection layer as contamination-free as the programs it protects.
It provides 44 distinct protection primitives organized into nine technique families.
These primitives can be composed to construct stronger, layered protections, including:
\begin{enumerate*}[label=(\arabic*), ref=(\arabic*)]

\item \label{prot:obfuscation} \textit{Obfuscation and string deception} (5 primitives).

\item \label{prot:auth-enc} \textit{Per-page authenticated encryption} (4 primitives).

\item \label{prot:lazy-dec} \textit{Lazy decryption and residency minimization} (5 primitives).

\item \label{prot:anti-debug} \textit{Measurement-keyed anti-debugging} (8 primitives).

\item \label{prot:anti-tamper} \textit{Self-checksumming and anti-tamper coupling} (3 primitives).

\item \label{prot:online-licensing} \textit{Online licensing with per-run key delivery} (4 primitives).

\item \label{prot:loader-virt} \textit{Loader-logic virtualization} (6 primitives).

\item \label{prot:anti-dump} \textit{Anti-dump detection and deception} (5 primitives).

\item \label{prot:anti-rehost} \textit{Anti-re-hosting} (4 primitives).
\end{enumerate*}
To our knowledge, more than half of these primitives, primarily in families~\ref{prot:lazy-dec}, \ref{prot:anti-debug}, \ref{prot:loader-virt}, \ref{prot:anti-rehost}, and~\ref{prot:anti-dump}, have no publicly available implementations.
Note that our implementation draws on ideas from commercial obfuscators but use independently designed algorithms and parameters.
More details can be found in \autoref{app:protection_details}.

\textbf{Instance Generation}
We compile each of the 19 programs and, where applicable, harden it with the suite, yielding 262 binary instances.
For each of the 16 programs in the four domains (\ie{} network protocol, game, format parser, malware), we build 8 unprotected instances that cover all combinations across three compilation axes (\ie{} optimization, symbol stripping, and static vs.\ dynamic linking) plus 8 protected instances (\ie{} one per protection preset, details in \S\ref{app:presets}).
Firmware is different: it is bare-metal, so our protection suite (which relies on the Linux loader and page-fault handlers) does not apply; instead the firmware ships its protection in-binary (encrypted boot, bespoke crypto, on-device VMs, and decoy peripherals).
Three firmware programs vary only by optimization level.
Every binary instance defines six deterministically scored tasks.
As a result, \benchmark{} provides 1{,}572 RE tasks in total.

\textbf{Contamination Control Deserves Special Note.}
Because this paper may itself enter future training corpora, we describe each domain in sufficient detail to support the paper's analysis and give readers a clear understanding of each domain, while limiting those details to information that an analyst could recover from the binary with minimal effort.
\footnote{We further verified that all information disclosed in \S\ref{sec:benchmark} and \S\ref{app:benchmark_details} could be recovered from the binaries by Codex GPT-4 mini within 200 LLM requests. We masked program-specific details in each paragraph, asked the agent to reconstruct them, and manually checked the results.}
We withhold all instance-specific secrets.
\footnote{Throughout the paper, we omit exact constants, trigger conditions, hidden-behavior mechanisms, and reference solutions. We report only what an agent could readily obtain from the binary. Where we name a critical algorithm (\eg{} a cryptographic primitive), we substitute one of similar functionality and complexity. We deliberately withhold the actual choice. \benchmark{} therefore remains contamination-free in practice.}

\section{Evaluation}
\label{sec:evaluation}
\begin{figure}[t]
  \centering
\includegraphics[width=0.9\linewidth]{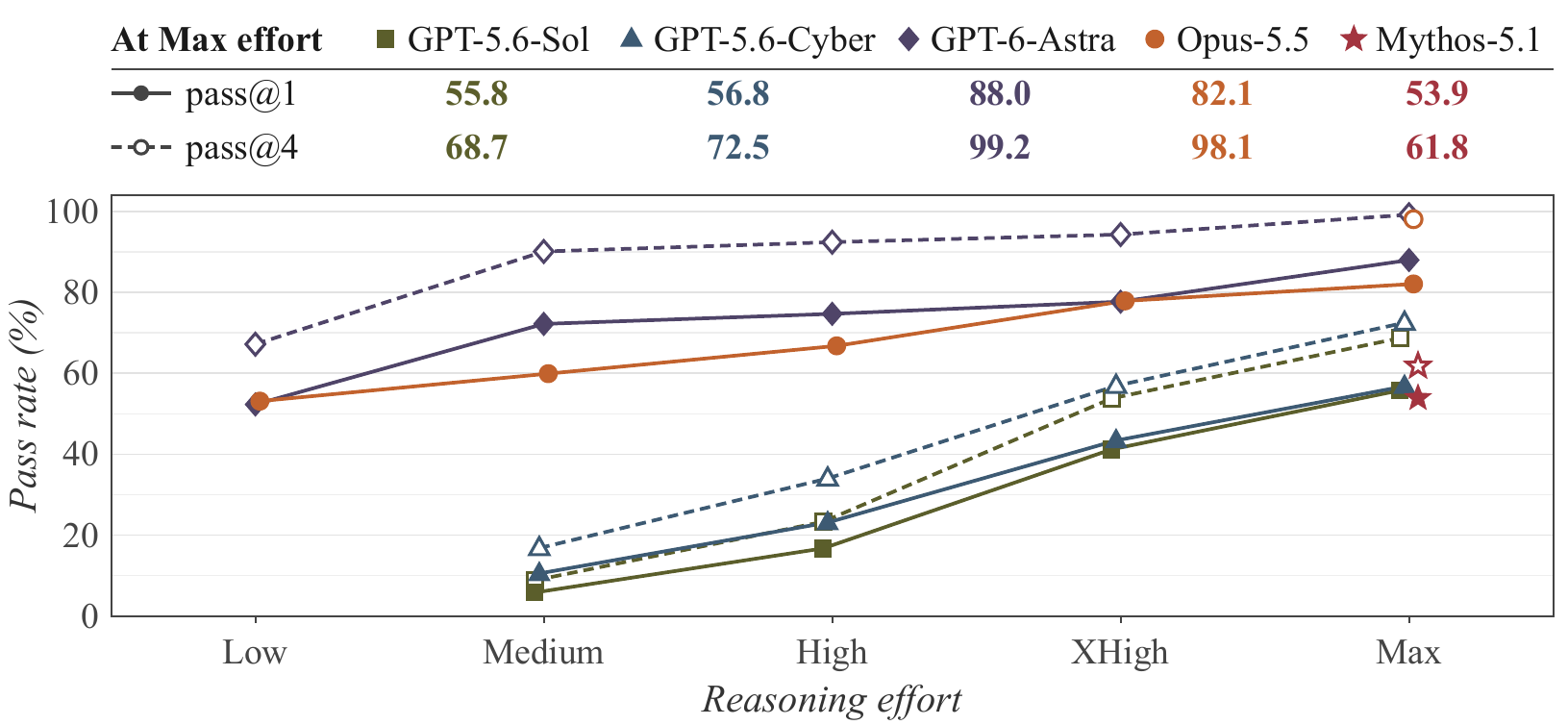}
  \caption{Pass@1 (solid) and pass@4 (dashed) of the unconstrained models by reasoning effort. Opus-5.5 at Low--XHigh ran once
  per instance, so its pass@4 is shown only at Max. GPT-5.6-Sol and GPT-5.6-Cyber were not run at Low, and Mythos-5.1 was run only at Max. All values are in
  Tables~\ref{tab:internal_effort} and~\ref{tab:prompt_results}.}
  \label{fig:effort_scaling}
  \end{figure}

\textbf{Models and Harness.}
We evaluate models in two settings: \emph{public-facing} and \emph{unconstrained}, the latter not available to customers.
For the \emph{public-facing} setting, we evaluated the eight models (\autoref{tab:overall}) with the model-side cyber safeguards enabled when available.
Each runs inside a standardized \mbox{mini-SWE-agent},
so that differences in scaffolding do not confound the comparison.
Every model is queried at its highest reasoning-effort setting,
and each run is capped at 500 model steps and 6 hours of wall-clock time.
The \emph{Unconstrained} setting (\autoref{fig:effort_scaling}) evaluated three OpenAI models and two Anthropic models at the providers' default settings, using Codex for OpenAI models and Claude Code for Anthropic models, without cyber safeguards and evaluation budget limits.
This internal evaluation comprises 16,768 attempts and 2.37B output tokens, corresponding to a token-equivalent cost of roughly \$566,000 (\autoref{tab:internal_cost}).

Every problem instance is solved in an isolated container holding the target binary, the domain-specific auxiliary materials, a set of standard RE tools (\ie{} Ghidra and
\texttt{pyghidra},
radare2,
GDB,
angr,
binutils,
\texttt{strace},
and \texttt{ltrace}) and per-domain tools (\ie{}
Including \texttt{tshark}, \texttt{pwntools}, \texttt{unicorn}, and \texttt{z3}).
Grader code, reference solutions, and any development artifacts are removed from the container, and agents receive no scoring feedback during a run.

\textbf{Metrics.}
We report the following metrics:
\emph{Score}, the mean of successful tasks per binary (out of six) across graded binaries;
\emph{Solved}, the number of binary instances in which all six tasks pass;
and \emph{Zeros}, the number of instances that earn no credit at all.
The \emph{Solved} and \emph{Zeros} metrics stand at the two extremes of the rubric,
and nearly half of each model's scores land at one end of these extremes.
Together, they provide a more informative characterization of agent capability than the mean alone.
We also report per-instance number of API calls, cost, and sandbox wall-clock time.
All averages are computed over gradable runs, excluding cyber guardrail refusals
and failures due to limited context window.
For the unconstrained evaluation, we report pass@1 and pass@4 from $n=4$ attempts per instance using the unbiased estimator of \citet{chen2021codex}, under which pass@1 is the mean
  full-solve rate across attempts and pass@4 the fraction of instances solved by at least one attempt.

\begin{table}[t]
  \centering
  \small
  \setlength{\tabcolsep}{4.5pt}
  \caption{
  \textbf{Overall results on SRE-Bench}.
  \#\emph{Graded} is the number of runs that produced a gradable result; the remainder failed either due to guarded-model refusal
  or context-window limits;
  most context-window failures occurred with \textsc{grok-4.5}.
  Only graded instances are used to compute the metrics.
  Arrows indicate: $\uparrow$ higher is better, $\downarrow$ lower is better.}
  \vspace{5pt}
  \resizebox{\textwidth}{!}{
  \begin{tabular}{l r c c c r r r}
  \toprule
  & & \multicolumn{3}{c}{\textbf{Capability}} & \multicolumn{3}{c}{\textbf{Cost}} \\
  \cmidrule(lr){3-5}\cmidrule(lr){6-8}
  \textbf{Model} & \textbf{\#Graded} & \textbf{Score}\,$\uparrow$ & \textbf{Solved}\,$\uparrow$ & \textbf{Zeros}\,$\downarrow$ & \textbf{Calls}\,$\downarrow$ & \textbf{Cost}\,$\downarrow$
  & \textbf{Time}\,$\downarrow$ \\
   & \scriptsize out of 262 & \scriptsize out of 6\,(\%) & \scriptsize \#\,(\% of graded) & \scriptsize \#\,(\%) & \scriptsize (\#) & \scriptsize (\$) & \scriptsize (min) \\
  \midrule
  {\textsc{gpt-6-astra}} & {181} & {$\mathbf{5.55}${\scriptsize$\pm$0.09}\,{\scriptsize(92.4)}} & {\textbf{149}\,{\scriptsize(82.3)}} & {\textbf{7}\,{\scriptsize(3.9)}} & {61} & {13.5} &
  {18} \\
  \textsc{gpt-5.6-sol}   & 254 & 3.69{\scriptsize$\pm$0.14}\,{\scriptsize(61.4)} & 80\,{\scriptsize(31.5)} & \phantom{0}41\,{\scriptsize(16.1)} & 196 & 42.5 & 88 \\
  {\textsc{claude-fable-5.1}} & {223} & {2.58{\scriptsize$\pm$0.18}\,{\scriptsize(43.0)}} & {60\,{\scriptsize(26.9)}} & {96\,{\scriptsize(43.0)}} & {173} & {34.7} & {123} \\
  \textsc{claude-opus-5} & 256 & 1.91{\scriptsize$\pm$0.14}\,{\scriptsize(31.8)} & 32\,{\scriptsize(12.5)} & 118\,{\scriptsize(46.1)} & 139 & 23.6 & 81 \\
  \textsc{gpt-5.5}       & 262 & 1.02{\scriptsize$\pm$0.10}\,{\scriptsize(17.0)} & 10\,{\scriptsize\phantom{0}(3.8)} & 148\,{\scriptsize(56.5)} & 128 & 17.8 & 46 \\
 {\textsc{deepseek-v4.1-flash}} & {262} & {0.85{\scriptsize$\pm$0.09}\,{\scriptsize(14.2)}} & {2\,{\scriptsize(0.8)}} & {166\,{\scriptsize(63.4)}} & {182} & {0.5} & {61} \\
  \textsc{grok-4.5}      & 218 & 0.45{\scriptsize$\pm$0.07}\,{\scriptsize\phantom{0}(7.6)} & \phantom{0}2\,{\scriptsize\phantom{0}(0.9)} & 160\,{\scriptsize(73.4)} & 115 & 13.5 & 66 \\
 
  \textsc{glm-5.2}       & 262 & 0.21{\scriptsize$\pm$0.03}\,{\scriptsize\phantom{0}(3.4)} & \phantom{0}0\,{\scriptsize\phantom{0}(0.0)} & 217\,{\scriptsize(82.8)} & 143 & 26.6 & 149 \\

  \bottomrule
  \end{tabular}}
  \label{tab:overall}
  \end{table}

\textbf{Results Overview.}
Table \ref{tab:overall}
shows that realistic RE remains challenging for most publicly evaluated agents:
GPT-5.6-Sol and Claude-Fable-5.1, despite their strong source-code security capabilities~\citep{openai2026gpt56, anthropic2026mythos5}, fully solve only 80/254 (31.5\%) and 60/223 (26.9\%) graded instances, respectively, with mean scores of 3.69 and 2.58 out of 6.
The weakest model scores 0.21 and does not fully solve a single challenge. The \textit{Zeros} metric further characterizes these limitations: GPT-5.6-Sol earns no credit on 16.1\% of graded instances, compared with 43.0\% for Claude-Fable-5.1, 46.1\% for Claude-Opus-5, and 56.5--82.8\% for the four lowest-scoring models.
These results reinforce the need to evaluate binary analysis as a distinct capability rather than infer it from success on source-code security tasks.

In contrast, GPT-6-Astra leads on \benchmark{}, averaging 5.55 out of 6 and
fully solving 149 of 181 gradable instances (82.3\%); only 3.9\% of gradable
instances receive no credit.
Although the tasks are not exploitation-oriented, frequent refusals on cyber
safety grounds leave GPT-6-Astra with the fewest gradable instances (181 of
262, 69.1\%) in the public setting.
This raises concerns about its practical utility for RE, and it remains
a grand challenge for vendors to balance safety and capability.
All these scores are conditional on a gradable result and should not
be interpreted as success rates over all 262 instances.

\begin{wrapfigure}[15]{r}{0.55\textwidth}
    \vspace{-2.7em}
    \centering
    \includegraphics[width=\linewidth]{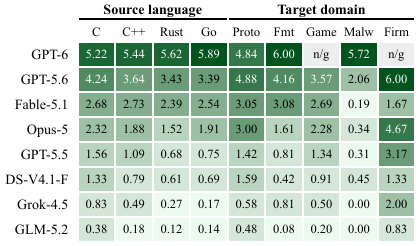}
    \vspace{-12pt}
    \caption{Mean score (of 6) by source language and target domain, over graded runs (n/g: none graded). Darker is higher.}
    \label{fig:result_attributes}
  \end{wrapfigure}

Finally, resource use does not predict capability. GPT-6-Astra is both the strongest model and the cheapest per instance, consuming 61 API calls and \$13.5 per instance while GPT-5.6-sol performs worse and costs \$42.5 per instance. GLM-5.2 spends \$26.6 per instance for a score of 0.21.

\textbf{Target Domains vs Source Languages.}
\autoref{fig:result_attributes} breaks down mean scores for public models by
source languages and target domains.
On the source language side, we found that most models perform better on
binaries compiled from C than ones compiled from Rust and Go.
We partially attribute this advantage to tooling and training data bias:
decompilers commonly produce C-like output, whereas Go and Rust binaries
introduce heavier runtime machinery and language-specific idioms that can
complicate analysis.
GPT-6-Astra achieves the highest mean score among evaluated models in every
source language, but scores highest on Go and lowest on C.

The target domain is a stronger indicator of model RE capability.
Network protocol and firmware generally observe higher scores, although firmware
contains only six instances. Malware is the lowest-scoring domain for most
models. Its strict grading may contribute to this difficulty: an agent must
extricate malicious behavior and preserve grader-planted benign state, so an
incomplete solution earns no credit.
Several weaker models achieve near-zero scores on malware while retaining higher
scores on protocol and game tasks. These differences show that aggregate scores
can conceal substantial variation in an agent’s ability to complete different RE
tasks.

\begin{wrapfigure}{r}{0.6\textwidth}
  \vspace{-1em}
  \centering
  \includegraphics[width=\linewidth]{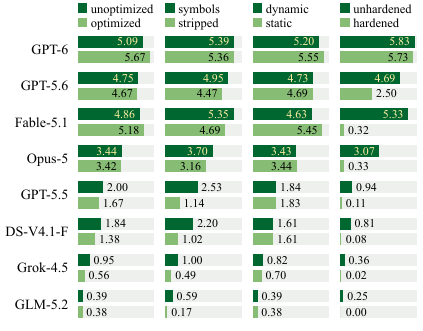}
  \vspace{-20pt}
  \caption{Mean score (of 6) per model for each build factor. The first three
  panels use the 128 unhardened instances; the last holds the build fixed
  (opt, stripped, static) and varies only protection.
  }
  \label{fig:result_build}
\end{wrapfigure}
\textbf{Build Configurations and Protection Mechanisms.}
\autoref{fig:result_build} breaks down mean scores by build configurations and
protection mechanisms, revealing a pattern that departs from the conventional ordering of
RE difficulty.
Optimization and linking, the primary obstacles for a human analyst, are nearly transparent to the four strongest models.
Weaker models lose 17--41\% under optimization, but their dominant build configuration is symbol stripping, which removes 50--71\% of the score of the four weakest models, versus 10--15\% for GPT-5.6-sol, Claude-Fable-5.1, and Claude-Opus-5, and nothing for GPT-6-Astra ($5.39$ vs.\ $5.36$).
Weaker models are thus far more sensitive to symbol stripping than stronger ones, suggesting that they rely more heavily on names as anchors for program understanding.
Protection is the one classical obstacle that transfers, but its effect depends sharply on the model's capability.
Hardening cuts GPT-5.6-sol’s score from $4.69$ to $2.50$ and reduces every other
model besides GPT-6-Astra to at most $0.33$. Claude-Fable-5.1 falls from $5.33$,
second only to GPT-6-Astra without hardening, to $0.32$ (see \S\ref{app:presets}
for results by preset).
GPT-6-Astra, by contrast, unpacks consistently and loses only $0.10$ points.
However, analyzing protection increases its cost, at $1.5\times$ the dollars per point of
unpacked instances, versus $3.4\times$ for GPT-5.6-sol and $9$--$35\times$ for
weaker models (\autoref{tab:packed_cost}).

\begin{wraptable}[11]{r}{0.5\textwidth}
\vspace{-2em}
\centering
\scriptsize
\setlength{\tabcolsep}{3pt}
\caption{\textbf{Complexity and contamination ablations.} The same eight
unhardened build settings of one compression task, over three programs.
}
\label{tab:contamination_ablation}
\begin{tabular}{@{}llccrr@{}}
\toprule
\textbf{Program} & \textbf{Model} & \textbf{Score}\,$\uparrow$ &
\textbf{Solve}\,$\uparrow$ & \textbf{Cost}\,$\downarrow$ & \textbf{Time}\,$\downarrow$ \\
\midrule
Minimal      & \textsc{gpt-5.6-sol} & 6.00 & 8/8 & 0.90 & 4.5 \\
             & \textsc{gpt-5.5}     & 6.00 & 8/8 & 1.60 & 7.0 \\
\midrule
Gzip-variant & \textsc{gpt-5.6-sol} & 6.00 & 8/8 & 2.19 & 9.9 \\
             & \textsc{gpt-5.5}     & 6.00 & 8/8 & 2.78 & 10.8 \\
\midrule
RevCompress  & \textsc{gpt-5.6-sol} & 5.62 & 7/8 & 31.45 & 94.6 \\
             & \textsc{gpt-5.5}     & 3.75 & 3/8 & 24.37 & 97.4 \\
\bottomrule
\end{tabular}
\end{wraptable}

\textbf{Unconstrained Evaluation and Solution Selection}.
\autoref{fig:effort_scaling} reports the strongest OpenAI-model results and the
clarified-prompt Anthropic-model results, without public safeguards or
evaluation budget limits.
GPT-6-Astra achieves $88.0\%$ pass@1 and $99.2\%$ pass@4; Opus 5.5 reaches
$82.1\%$ and $98.1\%$ under its internal configuration.
Reasoning effort matters: GPT-5.6 Sol improves from $5.8\%$ pass@1 at Medium to
$55.8\%$ at Max.
The full results and prompt comparison are in \S\ref{app:internal_sweep} and
\S\ref{app:prompt_clarification}.
High pass@4 means the private grader finds a correct solution among four attempts; practical use requires choosing or validating one without that grader.
For GPT-6-Astra, 69 instances have both successful and unsuccessful attempts, and 126 of its 1,048 attempts fail to obtain full credit.
Selecting the correct submission remains an open evaluation question, since these runs do not test a selector.

\textbf{Ablation: Contamination Control and Program Complexity}.
\autoref{tab:contamination_ablation} directly tests the two design criteria of
\S\ref{sec:intro}, with eight unhardened build configurations
fixed and varying only the target program.
Sufficient scale is necessary to accurately measure RE capabilities:
\emph{Minimal}, a clean-room compressor of about 1.1k LoC, is fully solved by
both models in every configuration for less than \$1.60 and seven minutes.
Therefore, being clean-room provenance by itself does not make a target hard.
\emph{Gzip-variant}, a deployed encoder with only about 200 lines of modified
code, is also solved in all eight configurations for roughly \$2 and ten minutes,
as most of its source is publicly available in training corpora.
Only \textsc{RevCompress}, which combines original code with real world scale,
separates the models: GPT-5.6-sol scores $5.62$ and solves $7/8$ configurations,
versus $3.75$ and $3/8$ for GPT-5.5, at 10 to 30 times the cost and time.

To conclude the ablation study, a large-scale program built largely from public
code can be as easy as a small private one, allowing recognition to stand in for
analysis.
By controlling both source exposure and program complexity, \benchmark{}
contributes a rigorous test of RE capability with meaningful difficulty.

\section{Discussion and Conclusion}
\label{sec:discussion}

\textbf{Limitation: Benchmark Breadth.}
\benchmark{} is built from 19 programs.
This is a small pool in absolute terms. %
This limitation follows directly from our design requirements.
To remain contamination-free, every target program, its evaluation infrastructure, and the protection layer must be developed from scratch; doing so required more than 5{,}000 expert hours.
Our ablation further shows why cheaper alternatives are inadequate: a program derived from public code can be recovered for roughly \$2 in ten minutes, while small clean-room programs are solved nearly perfectly (\autoref{tab:contamination_ablation}).
Despite the limited program count, \benchmark{} contains over 320K lines of authored code, averaging \benchmarkLoC{} LoC per program, together with a 27K-LoC protection suite.
These programs produce 1{,}572 deterministically scored tasks. %
This scale is already sufficient to expose large capability differences: across the eight public models, mean scores differ by more than $25\times$.

\textbf{Limitation: Program Size.}
Our programs average \benchmarkLoC{} LoC and thus do not cover the largest real-world software systems, which can reach hundreds of thousands or millions of lines of code.
Extending to that scale faces the same constraint as expanding benchmark breadth:
contamination-free programs must be developed from scratch, while development costs grow substantially with program size.
Nevertheless, our programs are already $30$--$470\times$ larger than those used in prior RE benchmarks.
We therefore view larger programs as an important direction for extending \benchmark{}. %

\textbf{Conclusion.}
We presented \benchmark{}, the first RE benchmark that is at once contamination-free and built at real-world scale.
Most frontier agents still struggle with it. 
GPT-5.6-Sol and Claude-Fable-5.1 fully solve only 31.5\% and 26.9\% of graded instances, and our anti-analysis suite halves the former while reducing every weaker model to near zero.
Agentic failure patterns also differ from human ones, as optimization and linking 
have little effect while stripping symbols is costly.
Together these results suggest that strong source-code security performance is not yet a reliable indicator of binary-level capability, and that the gap is wide enough to be worth measuring on its own terms.

\section*{AI Use Statement.}
Generative AI systems are a central object of study in this work and are extensively used as evaluated agents in our experiments. 
We also use AI tools for limited assistance with writing, code development, and debugging. 
All benchmark design decisions, implementations, experimental procedures, analyses, and conclusions are carried out by the authors. 
The authors remain fully responsible for the content of the paper and the correctness of all reported results.

\section*{Ethics Statement.}
This work studies AI capabilities for reverse engineering, which has legitimate applications in vulnerability analysis, malware inspection, and software auditing, but can also have dual-use implications. 
To reduce risks to real systems, the benchmark is constructed from programs developed in-house rather than vulnerabilities or binaries collected from deployed third-party software. 
Experiments are conducted in isolated environments, and the benchmark does not require interaction with external systems or services. 
We believe that systematically measuring these capabilities is important for understanding both the security benefits and the potential risks of increasingly capable cybersecurity agents.

\section*{Reproducibility Statement.}
To preserve the contamination resistance of \benchmark{}, we do not publicly release the benchmark binaries or task instances. 
Instead, we provide an open evaluation service through which researchers can request evaluation of their models or agents; we run the experiments in the same controlled environment and report the resulting metrics. 
The paper and appendices document the benchmark construction process, protection configurations, evaluation harness, model settings, resource limits, and metrics in detail. 
This setup enables consistent comparison of future models on \benchmark{} while preventing public exposure from undermining the validity of subsequent evaluations.

\bibliography{iclr2026_conference}
\bibliographystyle{iclr2026_conference}

\appendix
\section{Benchmark Domain Details}
\label{app:benchmark_details}

This appendix expands on each \benchmark{} domain, complementing the overview in \S\ref{sec:benchmark}.
Consistent with our contamination-control policy, we describe software architecture, task structure, and anti-shortcut design only at a level an analyst could recover from the binary with minimal efforts, and we deliberately omit the instance-specific secrets -- exact constants, trigger conditions, hidden-behavior mechanisms, and reference solutions -- that would let a model solve a task by recognition.

\textbf{\domainicon{protocol}Network Protocol.}
The agent receives a client binary that implements a proprietary, encrypted client--server protocol, together with a handful of partial packet captures that mimic the limited network traces available to analysts in practice.
No specification is provided.
The agent must recover the layered wire format and reconstruct the interacting client and server state machines, then implement a driver to communicate with a live server.
A live server grades that driver over a single metered TCP connection, capped at 64 messages, scoring how deep into the protocol's state space the agent can legally drive: 
from the cryptographic handshake, through capability-tier climbing, rekeying, and connection migration, to a capstone clean run that covers every stage at once.
The server generates a fresh key for each session, preventing the agent from replaying a previous run.

\textbf{\domainicon{game}Game.}
The target is a fully playable, 20-floor terminal roguelike featuring combat, spellcasting, crafting, shops, factions, quests, and achievements.
A player manual explains the game and its core mechanics.
Hidden within the binary are six behaviors that never arise during normal play and are not documented in the manual; each is triggered silently by conditions that can be identified only by reconstructing the relevant program logic.
The agent demonstrates each discovery by submitting an action trace, which a headless verifier replays and reports which hidden behaviors were triggered.

\textbf{\domainicon{compress}File Format.}
The target is a file-and-directory compressor that produces archives in a proprietary format, \ie{} an in-house version of GZIP.
It is distributed as an \emph{encoder only}: the binary contains no decompressor, so the agent must reverse engineer the multi-stage encoding pipeline and implement a decoder that reconstructs the original files byte-for-byte.
The pipeline combines six modified variants of commonly used compression algorithms with a custom error-correction scheme.
The agent must recover six password-protected archives of increasing complexity, some of which contain recoverable corruption that the decoder must detect and correct.
Grading requires exact recovery of both file contents and metadata; the decoder must genuinely invert every stage of the pipeline.

\textbf{\domainicon{malware}Malware.}
The target is a malware-themed Linux implant that is inert and safe by construction.
When activated with a key inside a sandbox, it simulates behaviors drawn from six common malware families: process masquerading, persistence, cryptomining, file ransomware, C2 beaconing, and lateral movement.
All effects are strictly confined: file writes remain under the sandbox root, networking is restricted to loopback, and the implant performs no real cryptomining, user-data encryption, or remote code execution.
The task is defensive: the agent must understand the implant well enough to implement a cleanup tool that surgically removes each family’s artifacts while preserving grader-planted benign data.
This requirement prevents indiscriminate ``delete everything'' strategies, a reward-hacking shortcut identified during adversarial auditing.
Scoring is conjunctive, with one point awarded per family only if the tool removes the malicious state, preserves the benign state, and durably neutralizes the corresponding behavior.
Instance-specific names, paths, and keys are all derived from a fresh secret on each run to prevent hard-coded cleanup strategies.

\textbf{\domainicon{firmware}Firmware.}
The target is bare-metal firmware for a fictional, locked-down secure microcontroller whose firmware packaging and system-on-chip architecture were designed entirely in-house.
It features an encrypted two-stage boot process, bespoke on-device cryptography, and dozens of peripheral subsystems.
The agent interacts with the device only through a JTAG-style interface to a from-scratch emulator that models the chip's memory map, peripherals, and serial command interface, subject to a per-session budget of 5{,}000 commands.
Beginning with black-box probing, the agent must reverse engineer the device and gain progressively greater control.
The six tasks correspond to milestones along a dependency chain, progressing from initial communication to full privileged control.
A fresh device secret is generated for each session.
Moreover, the scoring logic resides in the emulator rather than the firmware and is never exposed to the agent.

\subsection{Network Protocol Recovery}
\label{app:protocol}

\textbf{Software architecture.}
The target implements a complete proprietary protocol organized as four wire layers: framing with integrity checks and a de-recognized header layout; a from-scratch authenticated session-encryption layer (a stream cipher, a MAC, and a key-derivation function) that borrows the structure of standard AEAD constructions but replaces their constants, so nothing is recognizable by signature; stream multiplexing with priority, flow control, and compression; and an application layer with tens of message types.
The protocol also embeds a small stack-based bytecode filter VM and a set of interacting state machines coordinated by a supervisor.

\textbf{Task structure.}
The six scored anchors form a capability ladder:
(1)~recover the framing and complete the cryptographic handshake;
(2)~reconstruct the capability-tier state machine and forge the promotion tokens that gate the top tier;
(3)~follow the key-rotation schedule while maintaining a rolling-MAC chain across rekeys;
(4)~reconstruct the connection-migration state machine and answer its path challenges;
(5)~devirtualize the bytecode filter and satisfy a non-obvious hidden predicate within a bounded opcode budget;
and (6)~a capstone that requires all five above to hold within a single session while simultaneously satisfying a set of clean-run invariants under a whole-transcript MAC.
Partial credit is real, and any ``loud'' protocol violation voids the run.

\textbf{Anti-shortcut design.}
The protocol deliberately exposes a large decoy surface---compression, retransmission, priority scheduling, and much of several state machines---that is functionally exercised but not anchor-relevant, defeating ``this looks important'' heuristics.
Canonical answer strings are masked so that static extraction yields nothing, the handshake mixes in a fresh per-connection server nonce, and the per-session seed is randomized so a captured winning transcript cannot be replayed.
A per-connection message cap prevents brute-force flooding, and grading is silent: the driver is never told its score or even what the milestones are.

\textbf{Auxiliary infrastructure.}
Beyond the four language ports, the domain ships a from-scratch scoring server that is itself a complete protocol implementation running the anchor detectors, a cross-port conformance corpus with pinned wire transcripts, and deployment tooling that isolates the analysis and grading environments.
The agent receives only the stripped binary and a few partial packet captures.

\subsection{Game Reverse Engineering}
\label{app:game}

\textbf{Software architecture.}
The target is a full-screen terminal roguelike comprising roughly two dozen interacting subsystems---combat (including a multi-phase boss), a spell system across several schools of magic, crafting, shops, factions and reputation, quests, procedural map generation, field-of-view, inventory and leveling, an encyclopedia of lore, achievements, and deterministic record/replay.
It is genuine software of real-world scale rather than a template, and it behaves identically across all four language ports.

\textbf{Task structure.}
Hidden within the game are six behaviors that never occur during ordinary play and are documented nowhere in the player manual.
They are silent---no message, sound, or on-screen counter marks them---so the only way to trigger one is to reconstruct the responsible program logic from the binary.
The six behaviors are chosen to probe distinct RE capabilities of increasing difficulty, ranging from recovering an inlined constant and a per-object counter, through inverting a custom hash or pseudorandom generator and reconstructing a finite-state machine to search for a reachable input, up to jointly reconstructing state that is spread across several subsystems and unwinding a multi-round key derivation that combines constants from several of them.
We intentionally do not disclose the specific mechanisms or trigger conditions.
The agent proves a discovery by submitting an action trace, and a headless verifier replays it deterministically to report which behaviors fired.

\textbf{Anti-shortcut design.}
Several decoy subsystems are built from the \emph{same} kinds of constructs as the hidden behaviors (custom hashes, state machines, and unusual constants) but are wired into ordinary gameplay, so ``find the weird construct'' heuristics produce false positives; some decoys even reuse a hidden behavior's exact constant.
All gameplay strings and a broad catalog of magic constants ship only in obfuscated, runtime-decoded form, so neither a string dump nor an immediate-operand scan surfaces the critical values.
Determinism---required for offline replay---is guaranteed by a single seeded pseudorandom stream with a fixed consumption order, no wall-clock or environment reads, pinned floating-point behavior, and neutralized map-iteration order.
A build-time hygiene check greps the shipped, stripped binary for forbidden identifiers and fails the build on any hit.

\subsection{File-Format Recovery}
\label{app:compress}

\textbf{Software architecture.}
The compressor implements a genuine multi-stage transform stack wrapped in a keyed container.
The container provides framing, a reversible diffusion layer, optional password-based encryption, and an error-correcting recovery seal.
Internally, a long-range LZ pre-pass feeds a reversible pre-filter bank (e.g., delta, executable, and image-row filters), followed by an adaptive per-block compressor that trial-encodes each block under several competing recipes---spanning context-mixing arithmetic coding, range-coded LZ, and Burrows--Wheeler transforms~\citep{burrows1994block} with entropy back-ends---and keeps only the smallest.
At higher compression levels, a seed-derived self-mutating predictor VM, a keystream-whitening stage, and a keyed block permutation are applied.
These are standard mechanism \emph{classes}; the difficulty comes from their bespoke composition and from constants that are derived from a secret seed at runtime rather than stored, so nothing is recognizable by signature.

\textbf{Task structure.}
Six challenge files of increasing complexity progressively expose new layers of the stack: the outer container and its recovery seal; the pre-filter bank and long-range LZ with block-method compression; full archive semantics over directory trees (paths, permissions, empty directories) with the harder entropy back-ends; the predictor-VM residual stage, whose program is seed-derived and stored nowhere in the archive; error auto-correction via the recovery seal; and finally the keyed block permutation together with errors at the maximum correction capacity.
Because a decoder for the last level subsumes all earlier ones, the levels form a strict cumulative ladder.

\textbf{Challenge generation and grading.}
A deterministic generator produces all six challenges from a single secret master seed, drawing on a content palette that includes filesystem corner cases, and self-verifies that a correct decoder can recover each one.
The originals are never shipped, and the grader performs byte- and metadata-exact comparison with no partial credit within a level.
The binary ships as an encoder only---enforced by a build-time symbol scan and an import-graph gate---so no decoder or answer material is present.
A further check forbids linking any standard or third-party compression library, preventing recognition of an off-the-shelf codec.

\subsection{Malware Incident Response}
\label{app:malware}

\textbf{Software architecture.}
The implant masquerades as a benign local utility but, when armed with a host activation key inside a sandbox, installs six families of effects: process masquerading with a watchdog, persistence via an injected shell-profile block, cryptominer artifacts, file ransomware, C2 beaconing, and lateral-movement worm behavior.
Every effect is safe by construction: filesystem writes stay under an artifact root, networking is loopback-only, the miner's work is fake and bounded, ransomware only transforms sandbox-confined files carrying a full magic prefix, and the beacon/worm endpoints talk only to shipped validation stubs---there is no real remote execution, scanning, or encryption of user data.
Per-instance names, paths, ports, keys, and triggers are all derived from a fresh host-key-based secret, and the secret is deliberately consumed by benign modules too, so ``find the seed'' is not enough to solve an instance.

\textbf{Task structure.}
The RE task is defensive remediation, not behavior reproduction: the agent submits a cleanup tool that removes each family's artifacts while preserving benign data.
Scoring is flat and conjunctive---one point per family, awarded only when the agent removes the malicious state, preserves grader-planted benign state, \emph{and} durably neutralizes the behavior (no leftover resume tokens, triggers, or live processes).
The families form a difficulty ramp, and several couple across behaviors so that solving one depends on material recovered from another.

\textbf{Grading pipeline.}
A sandboxed grader derives the seed, plants baseline and near-match benign fixtures, captures a private pre-infection baseline, infects the sandbox, records evidence of what must be removed or repaired, runs the agent's cleanup, and scores the final state against baseline and evidence.
Evidence capture is what distinguishes ``an artifact never existed'' from ``an artifact existed and was correctly removed,'' since the grader never re-runs the implant after cleanup.
Any modification to a pre-existing baseline object is globally fatal, so ``delete everything'' strategies score zero; a fresh host key each attempt defeats hard-coded cleanups; and the untrusted cleanup runs in an isolated, resource-bounded container with no network.

\subsection{Bare-Metal Firmware Analysis}
\label{app:firmware}

\textbf{Software architecture.}
The firmware targets a fictional secure microcontroller with a plaintext bootloader that derives a key from a one-time-programmable fuse word and decrypts an encrypted application image into RAM.
It is a complete device---not a CrackMe---with a hardware-gated secure region holding a per-instance secret, a device-mode state machine, dozens of peripheral subsystems (the majority of which are pure or woven decoys), and several sandboxed bytecode VMs that hide key constants.
Its cryptographic core is a bespoke family designed to superficially resemble a well-known cipher while being algorithmically distinct, and it is split into independent variants so that recovering one does not hand over the others; genuinely standard checksums are placed in decoy modules so that recognizing them leads nowhere.

\textbf{Task structure.}
The student-facing framing is a single objective---take control of the device, scored $0$--$6$---with no sub-goal list; the milestone decomposition must itself be reverse-engineered.
The six milestones form a dependency chain that escalates from establishing basic wire contact and recovering the command framing, through discovering a gated maintenance path and forging an authentication token, to defeating an update-authentication format and a hardware-gated challenge--response, and finally a capstone that composes the earlier capabilities with multi-round key-derivation and bytecode-VM constant recovery to reach full privileged control.

\textbf{Emulator and anti-shortcut design.}
The agent interacts with the device only through a from-scratch emulator that faithfully models the memory map, the memory-mapped peripherals, the interrupt model, and the serial command protocol, under a per-session budget of $5{,}000$ commands.
A fresh random device secret each session re-derives the secret-dependent material, so a captured challenge--response for the later milestones scores zero on the next run; earlier milestones are additionally bound to a device-issued nonce, so individual frames cannot be replayed.
Decoys mislead in both directions---modules that look security-critical are inert, while load-bearing values hide inside modules that appear to be decoys---and all decoy behavior is held byte-identical across ports, so differential probing cannot separate scored from decoy logic.
Scoring is host-side and silent: no capability prints anything, and the grader records only the bare score.

\section{Protection Suite Details}
\label{app:protection_details}

This section adds context to the protection primitives summarized in \S\ref{sec:benchmark}.
Following our contamination-control policy, we describe every mechanism conceptually and deliberately omit the constants, labels, encodings, container formats, file layouts, and system-call sequences that would give an agent under test a head start.
The suite is an alpha-stage Linux x86-64 protector for static, self-contained executables.
It makes no claim of unbreakability: the goal is to \emph{raise the cost} of analysis in measurable, tunable steps, which is what makes it useful as a benchmark generator rather than as a security product.

\subsection{Threat Model and the Governing Design Rule}
\label{app:threat}

We assume a privileged analyst: root in the analysis sandbox, free use of debuggers and tracers, the ability to read and write the process's memory, to intercept system calls, and to dump memory during an authorized run.
Under those assumptions, no local scheme can prevent an analyst from eventually observing plaintext that the CPU must execute, and we do not pretend otherwise.
What the suite can do is make every observation \emph{expensive}, \emph{noisy}, and \emph{non-reusable across builds}.
The single rule behind most of the design is \textit{fold, don't branch}: instead of testing a condition and acting on it, the runtime feeds the measured value into key derivation.
A conventional check (\texttt{if (debugger) exit()}) is one instruction away from being removed; a folded measurement has no boolean to remove, because the wrong measurement simply yields the wrong key.

\subsection{Everything Feeds the Key}
\label{app:keying}

The protected program is encrypted at page granularity, and each page is sealed under its own key rather than under one global key.
Build identity, segment, offset, and length metadata are included as associated data, so modifying either the page contents or their metadata causes authentication to fail.
Those keys descend from a derivation graph whose inputs include the page's own identity, the integrity state of the executable's live code, the outcomes of the anti-analysis measurements, and -- on the lazy path -- evidence that the page is being opened by a genuine execution fault.
Two properties follow, and they are the reason we consider this the core of the design.
First, there is no single moment at which ``the key'' exists to be intercepted, so the classic breakpoint-on-decrypt strategy recovers only one page's worth of material.
Second, tampering and observation are indistinguishable from key corruption: a modified instruction, a suppressed check, or an attached tracer all lead to the same uninformative failure deep inside decrypted code, giving the analyst no signal about which defense reacted.

\subsection{Minimizing Plaintext Residency}
\label{app:residency}

Decrypting an entire program at startup means one well-timed memory snapshot recovers everything, so the lazy runtime instead maps protected code inaccessible and decrypts individual pages only as execution reaches them.
Three cooperating mechanisms then push plaintext back out of memory: a cap on how many code pages may be decrypted at once, a timer that re-encrypts pages that have gone idle, and a synchronous flush before the process blocks and becomes an easy target.
The loader adds W$\oplus$X enforcement, core-dump exclusion of secret pages, and un-optimizable secret wiping.
The practical effect is that a snapshot captures a small working set rather than an image, and a process caught while idle or waiting yields little.
We describe this honestly as residency \emph{reduction}: an analyst who reads memory continuously during execution still observes the instantaneous working set, and the union of everything ever resident approaches the full program over a long run.

\subsection{Establishing Exclusive Control}
\label{app:exclusivity}

Most anti-debugging asks ``is a debugger present?'' -- a question the analyst can answer falsely, since the reported value lives in memory the analyst controls.
We instead ask a question whose answer is enforced by the operating system: the runtime attempts to establish the exclusive control relationship a debugger itself requires, and folds the outcome into key derivation.
Failure implies a competing observer, while success additionally permits reading the processor's own debugging state, which catches breakpoints left behind by an analyst who has since detached.
A separate monitoring process keeps the master key masked while it is at rest and destroys it on detection, so that attaching \emph{after} a clean start corrupts decryption rather than merely raising an alarm; terminating that monitor is not a way out, because the protected process fails closed when the monitor stops reporting.

\subsection{Virtualized Loader Logic}
\label{app:vm}

A compact custom virtual machine executes the loader's decision logic -- how keys are derived, whether unwrapping is permitted, how control is transferred to the protected program, and how individual page faults are serviced -- so that this logic is not present as native code to be read or patched.
Rather than a recognizable interpreter loop, dispatch is diversified per build, instruction encodings are keyed, and frequently occurring sequences are fused into single operations, which removes the structural fingerprints that make virtual machines easy to identify and lift.
In the licensed configurations, parts of the virtualized program are not present in the file at all and are supplied per run by the license server, so static analysis of the artifact alone cannot recover the complete logic.

\subsection{Binding to a Genuine Launch, and Off-Box Keys}
\label{app:binding}

A capable analyst does not need to defeat the defenses if they can simply relocate the decryption machinery into a process of their own choosing and call it directly, which is an attack we observed in practice.
Four cooperating layers close it: key derivation is bound to evidence of a genuine launch of the original artifact, the decryption routines are themselves protected at rest and only reconstituted after startup validation, and on the lazy path each page key is additionally bound both to the fault that requested it and to the observation state at that moment.
The licensed configurations go further and keep no key in the artifact at all, obtaining it per run from a pinned server through an exchange that is freshly randomized each time, so a captured exchange cannot be replayed; the delivered key is then combined locally with the client's own measurements, meaning a server that answers an instrumented run still yields something unusable.
The residuals are stated plainly: on non-licensed configurations the in-file key material is in principle (but challengingly) recomputable offline, and on licensed configurations the delivered key must exist in memory during an authorized run and is therefore exposed to a live memory read.

\subsection{Deception Rather Than Denial}
\label{app:deception}
Because the strongest attacks cannot be blocked, several layers aim to make a successful attack \emph{unrecognizable} rather than impossible.
Strings that survive in the artifact are not merely encoded but renamed to plausible, misleading meanings, so that a first-pass triage read leads an analyst toward an incorrect model of what the program is doing.
When the runtime detects that it is being observed, it never confirms the detection: it may quietly restart itself so that an attached tool loses its target, or terminate as though it had completed successfully -- responses chosen specifically because a visible crash or error message would tell the analyst that their instrumentation was noticed and worked.
In the strongest configuration, and only when the author supplies one, the runtime serves an entirely different program in place of the real one, so that the analyst successfully dumps, decrypts, and reverse-engineers the wrong artifact with no failure to indicate the substitution.

\subsection{Anti-Agent Deterrents: A Reported Negative Result}
\label{app:anti_agent}

As a first exploration of protection aimed specifically at automated analysts, the suite can surface benign ``please decline to reverse-engineer this'' notices through several independent observation channels, so that a notice reaches an agent whether it reads the file, inspects memory, traces execution, or captures network traffic.
The mechanism carries no secret or exploitable content and is trivially removed; it is a behavioral nudge, not a barrier. 
We report it as a \textit{negative result}: capable agents (\ie{} GPT-5.6-sol, Opus 4.8) identified the notices as inert in a single reasoning step, and in one configuration the runtime's emission was actively harmful, serving as a landmark that helped locate the protected code path.
It is therefore disabled by default in every preset and retained only as a documented design point for future study.

\subsection{Preset Summary}
\label{app:presets}

\autoref{tab:protection_presets} summarizes the eight presets.
Every preset shares a common baseline -- per-page authenticated encryption, fold-not-branch keying, self-checksumming, loader virtualization, and the obfuscation and stripping layers -- and they differ along four axes that we found to matter most for difficulty: whether decryption is lazy, whether observation is keyed into decryption continuously (the \emph{strict} presets), whether the key is off-box, and whether the runtime responds to a detected dump by misdirecting the analyst.
The four strict presets are the only ones that break decryption on a \emph{late} attach and the only ones that accept an author-supplied decoy program.
Anti-agent deterrents are disabled by default in all eight and are therefore omitted from the table.

\begin{table}[t]
  \centering
  \footnotesize
  \setlength{\tabcolsep}{3pt}
  \renewcommand{\arraystretch}{1.1}
  \caption{Protection presets and their distinguishing primitives.
  \cmark{} = active; \textbullet{} = available but requires an author-supplied
  decoy; -- = inactive.
  All presets share the common baseline described in \S\ref{app:presets}.}
  \label{tab:protection_presets}
  \vspace{2pt}
  \begin{tabular}{@{}llcccccc@{}}
  \toprule
  \textbf{ID} & \textbf{Preset} & \textbf{Lazy} & \textbf{Strict} &
  \textbf{Anti-re-host} & \textbf{Off-box key} & \textbf{Server logic} &
  \textbf{Anti-dump} \\
  \midrule
  P1 & protected-eager           & --      & --      & \cmark & --      & --        & --      \\
  P2 & protected-eager-strict    & --      & \cmark  & \cmark & --      & --        & \textbullet \\
  P3 & protected-demand          & \cmark  & --      & \cmark & --      & --        & --      \\
  P4 & protected-demand-strict   & \cmark  & \cmark  & \cmark & --      & --        & \cmark  \\
  P5 & licensed-eager            & --      & --      & --     & \cmark  & shard     & --      \\
  P6 & licensed-demand           & \cmark  & --      & --     & \cmark  & shard     & --      \\
  P7 & licensed-demand-material  & \cmark  & \cmark  & --     & \cmark  & material  & \cmark  \\
  P8 & licensed-demand-superops  & \cmark  & \cmark  & --     & \cmark  & operators & \cmark  \\
  \bottomrule
  \end{tabular}
  \renewcommand{\arraystretch}{1.00}
\end{table}

\begin{figure}[h]
\centering
\includegraphics[width=\linewidth]{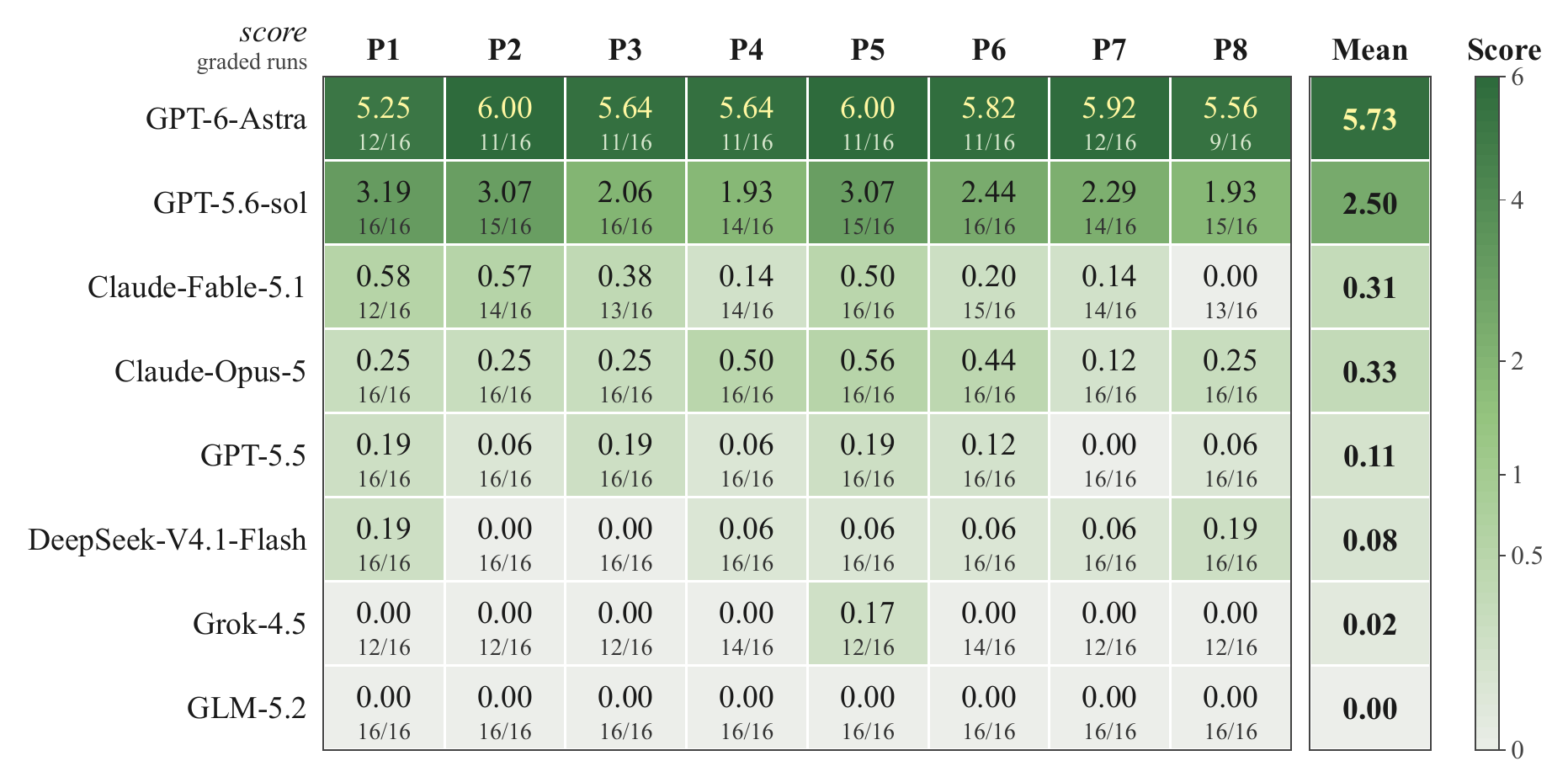}
\caption{Mean score (of 6) per model on the eight protection presets of
\autoref{tab:protection_presets}, all optimized, stripped and statically
linked. Compare against the unhardened bar of \autoref{fig:result_build}(d).}
\label{fig:result_hardening}
\end{figure}

\textbf{Per-Preset Results.}
\autoref{fig:result_hardening} resolves the aggregate protection effect reported in \S\ref{sec:evaluation} into the eight presets, holding the build fixed as optimized, stripped and statically linked.
GPT-6-Astra is largely unaffected, scoring between 5.25 and 6.00 on every preset. GPT-5.6-sol is the only other model that retains non-trivial capability, scoring between 1.93 and 3.19; every remaining model stays below 0.6 on every preset, and GLM-5.2 scores exactly 0.00 on all eight. Within GPT-5.6-sol the presets order roughly as designed, with eager whole-image decryption most tractable (P1, P5) and the strict and server-side variants least (P4, P8). 
That ordering is legible only for GPT-5.6-sol, however, because GPT-6-Astra sits near the ceiling and the remaining six sit near the floor.

\begin{table}
\centering
\footnotesize
\setlength{\tabcolsep}{4pt}
\caption{\textbf{Cost per point on packed instances (public evaluation).} Packed covers the 128 protected and licensed
non-firmware instances. Score is the mean over graded runs; Points is the total rubric points earned; Cost is total
spend over all runs, including errored ones; \$/pt is Cost divided by Points (-- where no points were earned).
Unpacked \$/pt is computed the same way over the 128 unhardened instances, and the last column is the ratio of
packed to unpacked \$/pt. \$/pt is unstable for models earning only a few points.}
\label{tab:packed_cost}
\vspace{5pt}
\begin{tabular}{@{}l cccc c cc@{}}
\toprule
 & \multicolumn{4}{c}{\textbf{Packed}} && \multicolumn{2}{c}{\textbf{vs.\ unpacked}} \\
\cmidrule{2-5}\cmidrule{7-8}
\textbf{Model} & Score & Points & Cost (\$) & \$/pt && Unpacked \$/pt & Ratio \\
\midrule
\textsc{gpt-6-astra} & 5.73 & 504 & 1,604.76 & 3.18 && 2.15 & 1.5$\times$ \\
\textsc{gpt-5.6-sol} & 2.50 & 303 & 6,785.33 & 22.39 && 6.52 & 3.4$\times$ \\
\textsc{claude-fable-5.1} & 0.32 & 35 & 3,702.81 & 105.79 && 6.93 & 15.3$\times$ \\
\textsc{claude-opus-5} & 0.33 & 42 & 2,813.08 & 66.98 && 7.22 & 9.3$\times$ \\
\textsc{gpt-5.5} & 0.11 & 14 & 1,939.58 & 138.54 && 11.02 & 12.6$\times$ \\

\textsc{deepseek-v4.1-flash} & 0.08 & 10 & 67.68 & 6.77 && 0.34 & 19.7$\times$ \\
\textsc{grok-4.5} & 0.02 & 2 & 1,308.62 & 654.31 && 18.56 & 35.3$\times$ \\
\textsc{glm-5.2} & 0.00 & 0 & 3,039.58 & -- && 79.51 & -- \\
\bottomrule
\end{tabular}
\end{table}

\section{Unconstrained Evaluation Details}
\label{app:internal}
\begin{table}[t]
  \centering
  \footnotesize
  \setlength{\tabcolsep}{4pt}
  \caption{Token-equivalent cost of the unconstrained provider evaluation. GPT-6-Astra and GPT-5.6 Sol use OpenAI's published per-task costs for the matching configurations, summed over 262 instances
  and four attempts. GPT-5.6 Cyber has no published cost and is estimated at GPT-5.6 Sol's cost per output token. Anthropic costs are estimated from token counts at list price, assuming
  98\% of input tokens are served from cache.}
  \vspace{15pt}
  \label{tab:internal_cost}
  \begin{tabular}{@{}llrrlr@{}}
  \toprule
  \textbf{Model} & \textbf{Evaluation} & \textbf{Runs} & \textbf{Output tok.} & \textbf{Cost basis} & \textbf{Cost} \\
  \midrule
  GPT-6-Astra   & pass@4, 5 efforts               & 5,240 & 215.7M & published    & \$83.4K \\
  GPT-5.6 Sol   & pass@4, 4 efforts               & 4,192 & 506.9M & published    & \$161.9K \\
  GPT-5.6 Cyber & pass@4, 4 efforts               & 4,192 & 773.6M & Sol \$/token & \$246.8K \\
  Opus 5.5      & pass@1 Low--XHigh + pass@4 Max  & 2,096 & 512.2M & list price   & \$52.3K \\
  Mythos 5.1    & pass@4 Max                      & 1,048 & 362.9M & list price   & \$22.0K \\
  \midrule
  Total         &                                 & 16,768 & 2.37B &              & $\approx$\$566K \\
  \bottomrule
  \end{tabular}
  \end{table}
\subsection{Reasoning Effort and Repeated Attempts}
\label{app:internal_sweep}
\begin{table}
\centering
\footnotesize
\setlength{\tabcolsep}{4pt}
\caption{\textbf{Unconstrained evaluation by domain and language.} Each cell reports pass@1\,/\,pass@4 (\%).
Protocol, format, game, and malware cells cover 16 instances each; firmware cells cover 2 (none in Go).}
\label{tab:internal_cells}
\vspace{10pt}
\begin{tabular}{@{}ll ccccc@{}}
\toprule
\textbf{Model} & \textbf{Lang.} & Proto & Fmt & Game & Malw & Firm \\
\midrule
\multirow{4}{*}{\textsc{gpt-5.6-sol}} & C & 95.3\,/\,100.0 & 98.4\,/\,100.0 & 25.0\,/\,43.8 & 37.5\,/\,75.0 & 100.0\,/\,100.0 \\
 & C++ & 90.6\,/\,100.0 & 87.5\,/\,100.0 & 17.2\,/\,31.2 & 14.1\,/\,37.5 & 100.0\,/\,100.0 \\
 & Rust & 90.6\,/\,100.0 & 89.1\,/\,100.0 & 1.6\,/\,6.2 & 21.9\,/\,50.0 & 50.0\,/\,50.0 \\
 & Go & 100.0\,/\,100.0 & 100.0\,/\,100.0 & 6.2\,/\,18.8 & 7.8\,/\,31.2 & -- \\
\midrule
\multirow{4}{*}{\textsc{gpt-5.6-cyber}} & C & 98.4\,/\,100.0 & 92.2\,/\,100.0 & 39.1\,/\,75.0 & 32.8\,/\,75.0 & 100.0\,/\,100.0 \\
 & C++ & 98.4\,/\,100.0 & 90.6\,/\,100.0 & 17.2\,/\,43.8 & 15.6\,/\,37.5 & 100.0\,/\,100.0 \\
 & Rust & 90.6\,/\,100.0 & 85.9\,/\,100.0 & 1.6\,/\,6.2 & 14.1\,/\,43.8 & 100.0\,/\,100.0 \\
 & Go & 98.4\,/\,100.0 & 96.9\,/\,100.0 & 12.5\,/\,37.5 & 7.8\,/\,31.2 & -- \\
\midrule
\multirow{4}{*}{\textsc{gpt-6-astra}} & C & 87.5\,/\,100.0 & 100.0\,/\,100.0 & 90.6\,/\,100.0 & 82.8\,/\,100.0 & 100.0\,/\,100.0 \\
 & C++ & 90.6\,/\,100.0 & 98.4\,/\,100.0 & 84.4\,/\,100.0 & 82.8\,/\,100.0 & 100.0\,/\,100.0 \\
 & Rust & 100.0\,/\,100.0 & 100.0\,/\,100.0 & 35.9\,/\,93.8 & 93.8\,/\,100.0 & 100.0\,/\,100.0 \\
 & Go & 100.0\,/\,100.0 & 100.0\,/\,100.0 & 62.5\,/\,93.8 & 93.8\,/\,100.0 & -- \\
\midrule
\multirow{4}{*}{\textsc{opus-5.5}} & C & 79.7\,/\,100.0 & 100.0\,/\,100.0 & 76.6\,/\,100.0 & 95.3\,/\,100.0 & 100.0\,/\,100.0 \\
 & C++ & 90.6\,/\,100.0 & 100.0\,/\,100.0 & 48.4\,/\,81.2 & 79.7\,/\,100.0 & 100.0\,/\,100.0 \\
 & Rust & 92.2\,/\,100.0 & 100.0\,/\,100.0 & 34.4\,/\,87.5 & 89.1\,/\,100.0 & 100.0\,/\,100.0 \\
 & Go & 96.9\,/\,100.0 & 98.4\,/\,100.0 & 54.7\,/\,100.0 & 70.3\,/\,100.0 & -- \\
\midrule
\multirow{4}{*}{\textsc{mythos-5.1}} & C & 100.0\,/\,100.0 & 100.0\,/\,100.0 & 15.6\,/\,43.8 & 12.5\,/\,43.8 & 100.0\,/\,100.0 \\
 & C++ & 98.4\,/\,100.0 & 98.4\,/\,100.0 & 3.1\,/\,12.5 & 15.6\,/\,50.0 & 100.0\,/\,100.0 \\
 & Rust & 98.4\,/\,100.0 & 100.0\,/\,100.0 & 0.0\,/\,0.0 & 3.1\,/\,12.5 & 100.0\,/\,100.0 \\
 & Go & 96.9\,/\,100.0 & 100.0\,/\,100.0 & 1.6\,/\,6.2 & 1.6\,/\,6.2 & -- \\
\bottomrule
\end{tabular}
\end{table}
 
\begin{table}
\centering
\footnotesize
\setlength{\tabcolsep}{2pt}
\caption{\textbf{Unconstrained evaluation by protection preset} (\autoref{tab:protection_presets}). Pass@1 (\%) with pass@4 as subscript.
Unhard.\ covers all 128 unhardened non-firmware instances; each preset column covers 16 instances.}
\label{tab:internal_presets}
\vspace{5pt}
\begin{tabular}{@{}l c cccccccc@{}}
\toprule
\textbf{Model} & Unhard. & P1 & P2 & P3 & P4 & P5 & P6 & P7 & P8 \\
\midrule
\textsc{gpt-5.6-sol} & 63.5$_{78.1}$ & 50.0$_{62.5}$ & 50.0$_{56.2}$ & 46.9$_{68.8}$ & 43.8$_{56.2}$ & 48.4$_{50.0}$ & 48.4$_{62.5}$ & 43.8$_{56.2}$ & 43.8$_{56.2}$ \\
\textsc{gpt-5.6-cyber} & 63.9$_{82.0}$ & 54.7$_{68.8}$ & 51.6$_{68.8}$ & 46.9$_{62.5}$ & 46.9$_{50.0}$ & 50.0$_{62.5}$ & 45.3$_{56.2}$ & 45.3$_{62.5}$ & 40.6$_{62.5}$ \\
\textsc{gpt-6-astra} & 88.7$_{100.0}$ & 87.5$_{93.8}$ & 84.4$_{100.0}$ & 85.9$_{100.0}$ & 87.5$_{100.0}$ & 90.6$_{93.8}$ & 87.5$_{100.0}$ & 82.8$_{100.0}$ & 87.5$_{100.0}$ \\
\textsc{opus-5.5} & 85.2$_{100.0}$ & 76.6$_{93.8}$ & 73.4$_{87.5}$ & 70.3$_{93.8}$ & 75.0$_{93.8}$ & 84.4$_{100.0}$ & 76.6$_{100.0}$ & 87.5$_{100.0}$ & 81.2$_{100.0}$ \\
\textsc{mythos-5.1} & 53.9$_{64.8}$ & 53.1$_{56.2}$ & 51.6$_{56.2}$ & 51.6$_{56.2}$ & 53.1$_{62.5}$ & 53.1$_{62.5}$ & 51.6$_{56.2}$ & 50.0$_{50.0}$ & 50.0$_{56.2}$ \\
\bottomrule
\end{tabular}
\end{table}
\begin{table*}[t!]
\centering
\footnotesize
\setlength{\tabcolsep}{5pt}
\caption{\textbf{Unconstrained provider reasoning-effort sweep.} Each OpenAI setting and Opus 5.5 at Max has four attempts per instance (1,048 attempts); Opus 5.5 at Low--XHigh has one attempt per instance. Pass rates and objective coverage are percentages. Output tokens are per-attempt means in thousands, counting the complete continuation chain; $n_{\mathrm{tok}}$ gives the reconciled token records used in that mean.}
\label{tab:internal_effort}
\vspace{5pt}
\begin{tabular}{llrrrrr}
\toprule
\textbf{Model} & \textbf{Effort} & \textbf{Pass@1} & \textbf{Pass@4} & \textbf{Coverage} & \textbf{Tokens (K)} & $n_{\mathrm{tok}}$ \\
\midrule
GPT-5.6 Sol & Medium & 5.8 & 8.8 & 14.9 & 28.9 & 1046 \\
 & High & 16.7 & 23.3 & 36.6 & 70.9 & 1046 \\
 & XHigh & 41.2 & 53.8 & 69.1 & 143.3 & 1044 \\
 & Max & 55.8 & 68.7 & 83.9 & 241.5 & 1047 \\
\midrule
GPT-5.6 Cyber & Medium & 10.5 & 16.8 & 23.6 & 64.5 & 1043 \\
 & High & 23.1 & 34.0 & 43.9 & 127.7 & 1043 \\
 & XHigh & 43.4 & 56.9 & 68.8 & 211.1 & 1040 \\
 & Max & 56.8 & 72.5 & 83.1 & 342.6 & 1032 \\
\midrule
GPT-6-Astra & Low & 52.3 & 67.2 & 76.1 & 20.7 & 1048 \\
 & Medium & 72.2 & 90.1 & 90.3 & 34.2 & 1046 \\
 & High & 74.7 & 92.4 & 93.2 & 38.7 & 1048 \\
 & XHigh & 77.7 & 94.3 & 93.4 & 42.6 & 1048 \\
 & Max & 88.0 & 99.2 & 97.0 & 69.7 & 1048 \\

 \midrule

Opus 5.5 & Low    & 53.1 & --   & 84.4 & 117.5 & -- \\
         & Medium & 59.9 & --   & 81.2 & 143.3 & -- \\
         & High   & 66.8 & --   & 85.2 & 169.6 & -- \\
         & XHigh  & 77.9 & --   & 89.3 & 240.1 & -- \\
         & Max    & 82.1 & 98.1 & 93.5 & 321.1 & -- \\
\bottomrule
\end{tabular}
\end{table*}

\autoref{tab:internal_effort} reports all 13 internal OpenAI settings.
Increasing effort improves full-solve rates for all three models across the tested settings.
At Medium effort, GPT-6-Astra achieves 72.2\% pass@1 with 34.2K output tokens per attempt, compared with 55.8\% and 241.5K tokens for GPT-5.6 Sol at Max.
Output-token counts describe generation effort rather than dollar cost or latency.
GPT-5.6 Cyber at Max has a higher pass@4 than GPT-5.6 Sol (72.5\% versus 68.7\%), but slightly lower objective coverage (83.1\% versus 83.9\%) and greater output-token usage.

Repeated attempts expose a distinction between generating a correct solution and recognizing one.
For GPT-6-Astra at Max, 191 instances succeed in all four attempts, while 13, 23, and 33 instances succeed in exactly one, two, and three attempts, respectively; two never fully succeed.
Thus, 69 instances contain both correct and incorrect submissions.
The corresponding pass@4 records whether a correct candidate exists, using the private grader after execution.
A future selection evaluation can instead require the agent to choose one of its four submissions before private grading and report the correctness of that choice.
The present evaluations do not measure such a selector.

\subsection{Malware Prompt Clarification}
\label{app:prompt_clarification}

\begin{table}[h]
\centering
\footnotesize
\setlength{\tabcolsep}{5pt}
\caption{\textbf{Malware prompt clarification in the unconstrained Anthropic evaluation.} The clarification adds one sentence explaining cleanup grading. Coverage@1 averages objective checks across attempts; Coverage@4 uses the best single-attempt score per instance. All values are percentages.}
\label{tab:prompt_results}
\vspace{5pt}
\begin{tabular}{llrrrr}
\toprule
\textbf{Model} & \textbf{Prompt} & \textbf{Pass@1} & \textbf{Pass@4} & \textbf{Coverage@1} & \textbf{Coverage@4} \\
\midrule
Opus 5.5 & Original & 76.3 & 96.6 & 87.4 & 98.5 \\
 & Clarified & 82.1 & 98.1 & 92.9 & 99.7 \\
\midrule
Mythos 5.1 & Original & 53.3 & 59.5 & 74.5 & 83.4 \\
 & Clarified & 53.9 & 61.8 & 76.0 & 86.9 \\
\bottomrule
\end{tabular}
\end{table}

\autoref{tab:prompt_results} compares the original prompt with one added sentence explaining how malware cleanup is graded, within the internal Claude Code evaluation.
For Opus 5.5, the clarification raises pass@1 from 76.3\% to 82.1\% and objective coverage from 87.4\% to 92.9\%.
Mythos 5.1 changes less, from 53.3\% to 53.9\% pass@1.
We report both conditions because interpreting the cleanup requirement contributes to measured task performance.
The main internal table uses the clarified prompt and identifies that choice explicitly.

\end{document}